\documentclass{moriond}

\def\be{\begin{equation}}
\def\ee{\end{equation}}
\def\bea{\begin{eqnarray}}
\def\eea{\end{eqnarray}}

\newcommand{\Photo}{\includegraphics[height=35mm]{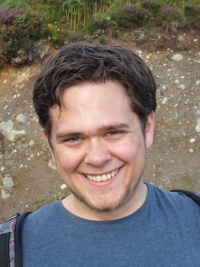}}

\newcommand{\pPb}{\textnormal{p--Pb}}

\newcommand{\snn}{{\sqrt{s_\mathrm{NN}}}}

\newcommand{\pT}{\ensuremath{p_\mathrm{T}}}

\begin{document}
\vspace*{4cm}
\title{Latest ALICE results of photon and jet measurements}

\author{ R\"udiger Haake }

\address{CERN,\\
CH-1211 Geneva 23, Switzerland}

\maketitle\abstracts{Highly energetic jets and photons are complementary probes for the kinematics and the topology of nuclear collisions.
Jets are collimated sprays of charged and neutral particles, which are produced in the fragmentation of hard scattered partons in an early stage of the collision. While traversing the medium formed in nuclear collisions, they lose energy and therefore carry information about the interaction of partons with the medium. The jet substructure is particularly interesting to learn about in-medium modification of the jets and several observables exists to probe it.
In contrast to jets, photons are created in all collision stages. There are prompt photons from the initial collision, thermal photons produced in the medium, and decay- and fragmentation photons from later collision stages. Photons escape the medium essentially unaffected after their creation.\\
This article presents recent ALICE results on jet substructure and direct photon measurements in pp, p--Pb and Pb--Pb collisions.}

\section{Introduction}

Jets can conceptually be described as the final state produced in a hard partonic scattering. Therefore, jets are an excellent tool to access a very early stage of a heavy-ion collision. The jet constituents represent the final state remnants of the fragmented partons that were scattered in the reaction. While all the detected particles have been created in a non-perturbative process (i.e. by hadronization), ideally, jets represent the kinematic properties of the originating partons. Thus, jets are mainly determined by perturbative processes due to the high momentum transfer and the cross sections can be calculated with pQCD. This conceptual definition is descriptive and very simple, the technical analysis of those objects is complicated though.\\

Since photons do not interact strongly, they essentially pass through the medium without being affected. While the jets carry information about the medium due to their in-medium energy loss and modification, photons retain the properties they had at their production. Direct photons, which are all photons except for decay photons, allow to test pQCD and to constrain PDFs. Thermal photons, which are produced by the medium, carry information about bulk properties like the temperature.

\section{Experimental details}
\label{sec:Details}

The data were recorded with ALICE, the dedicated heavy-ion experiment at the LHC studying properties of the quark-gluon plasma and the QCD phase diagram in general. The detector is designed as a general-purpose heavy-ion detector\,\cite{ALICE2008} to measure and identify hadrons, leptons, and also photons down to very low transverse momenta.

Charged jets have been measured in pp, p--Pb, and Pb--Pb collisions, using mainly data from the TPC\,\cite{ALICE2010b}, which is a time projection chamber, and the ITS\,\cite{ALICE2010} -- a six-layered silicon detector. The signals in these detectors are used to form charged-particle tracks that serve as the basic ingredient to jet reconstruction. By adding clusters measured by the EMCal\,\cite{Emcal}, an electromagnetic sampling calorimeter, full jets can be reconstructed with charged and neutral particles.
To measure jets, the anti-$k_\mathrm{T}$ algorithm\,\cite{Cacciari2008} implemented in FastJet\,\cite{Cacciari2006} is used in the present analyses. The track selection for those particles are chosen in order to obtain a uniform charged track distribution in the full $\eta-\phi$ plane for tracks with $\pT > 150 \mbox{ MeV/}c$ within pseudorapidity $|\eta| < 0.9$.

The mean background density is calculated on an event-by-event basis, using the median of clusters found by the $k_\mathrm{T}$ algorithm, leaving out the two highest-$\pT$ clusters. In the $\pPb$ system, the particle occupancy is also taken into account.\,\cite{KTBackgroundCMS}
Depending on the observable, this density is corrected with the area-derivatives method\,\cite{ADmethod}, constituent subtraction method\,\cite{CSmethod}, or jet-by-jet based on area.
Residual background fluctuations and detector effects are corrected for in a two-dimensional unfolding procedure.\\

Photons can be measured by three approaches. On the one hand, they can be directly measured by two calorimeters in the ALICE central barrel: The EMCal mentioned above and the homogeneous calorimeter PHOS, which consists of highly segmented PbWO${}_4$ crystals.
In addition, photons can also be reconstructed from charged tracks using the Photon Conversion Method (PCM). Here the secondary vertices of $e^{+}e^{-}$ pairs from photons which converted in the material of ALICE are reconstructed.

A statistical measurement of the direct photon signal can be achieved by subtracting the decay photons from all known hadronic sources from the inclusive photon sample. To extract the photon excess and to study its significance, we use the double ratio $R_\gamma = \frac{\gamma_\mathrm{inc}}{\gamma_\mathrm{dec}} = \frac{\gamma_\mathrm{inc}}{\pi^0} / \frac{\gamma_\mathrm{dec}}{\pi^0}$, where correlated systematic errors cancel. Here $\gamma_\mathrm{inc}$ and $\gamma_\mathrm{dec}$ are the inclusive and decay photon spectra and $\pi^0$ is a parametrization of the measured $\pi^0$ spectrum.\,\cite{DPRatio}

Isolated photons allow a very precise measurement of direct photons (excluding fragmentation photons) using an isolation criterion. The criterion is applied on the total energy in a cone radius $R=0.4$ around the photon candidates.\,\cite{IsolationCut}

\begin{figure}[htp]
\includegraphics[width=0.45\textwidth]{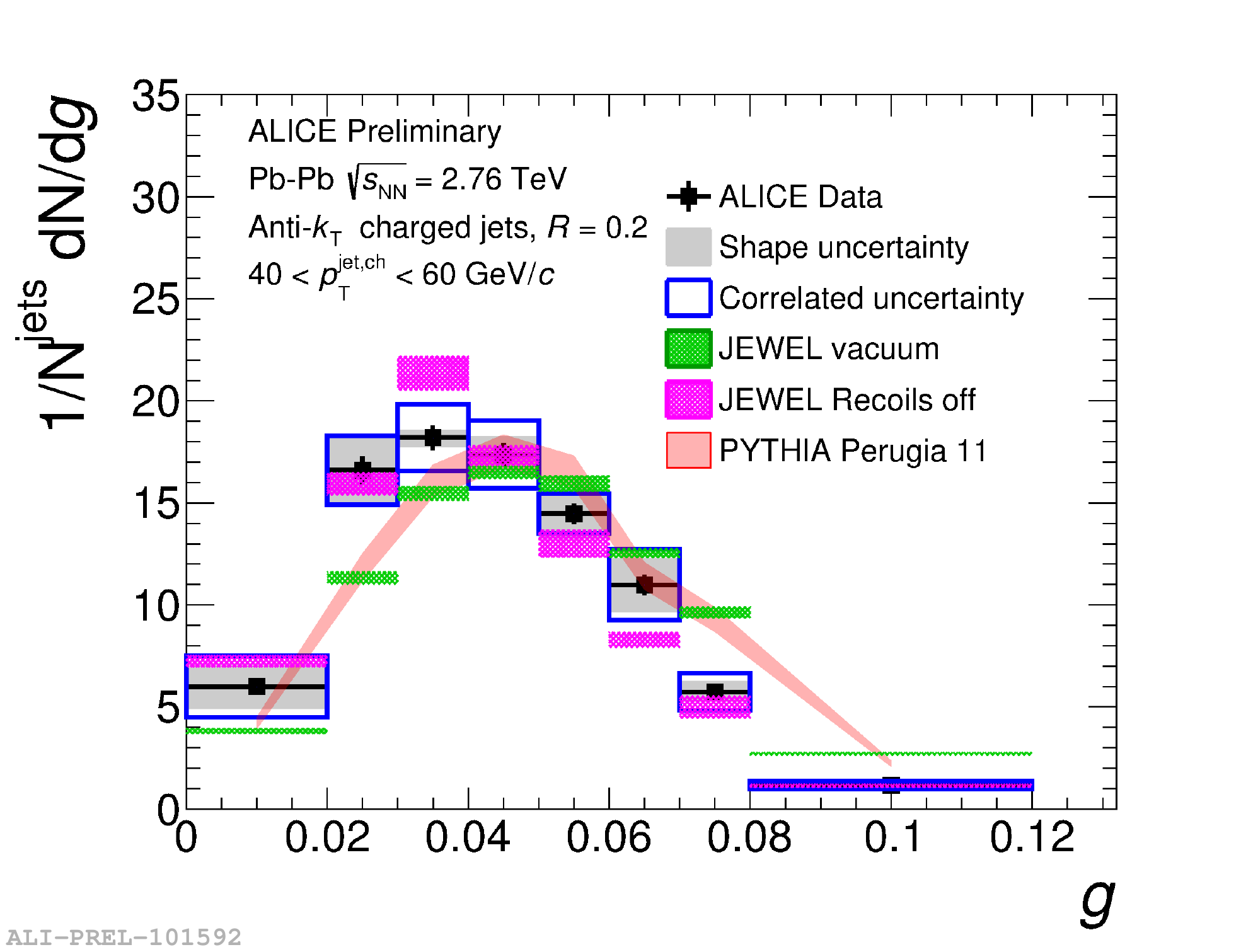}
\includegraphics[width=0.45\textwidth]{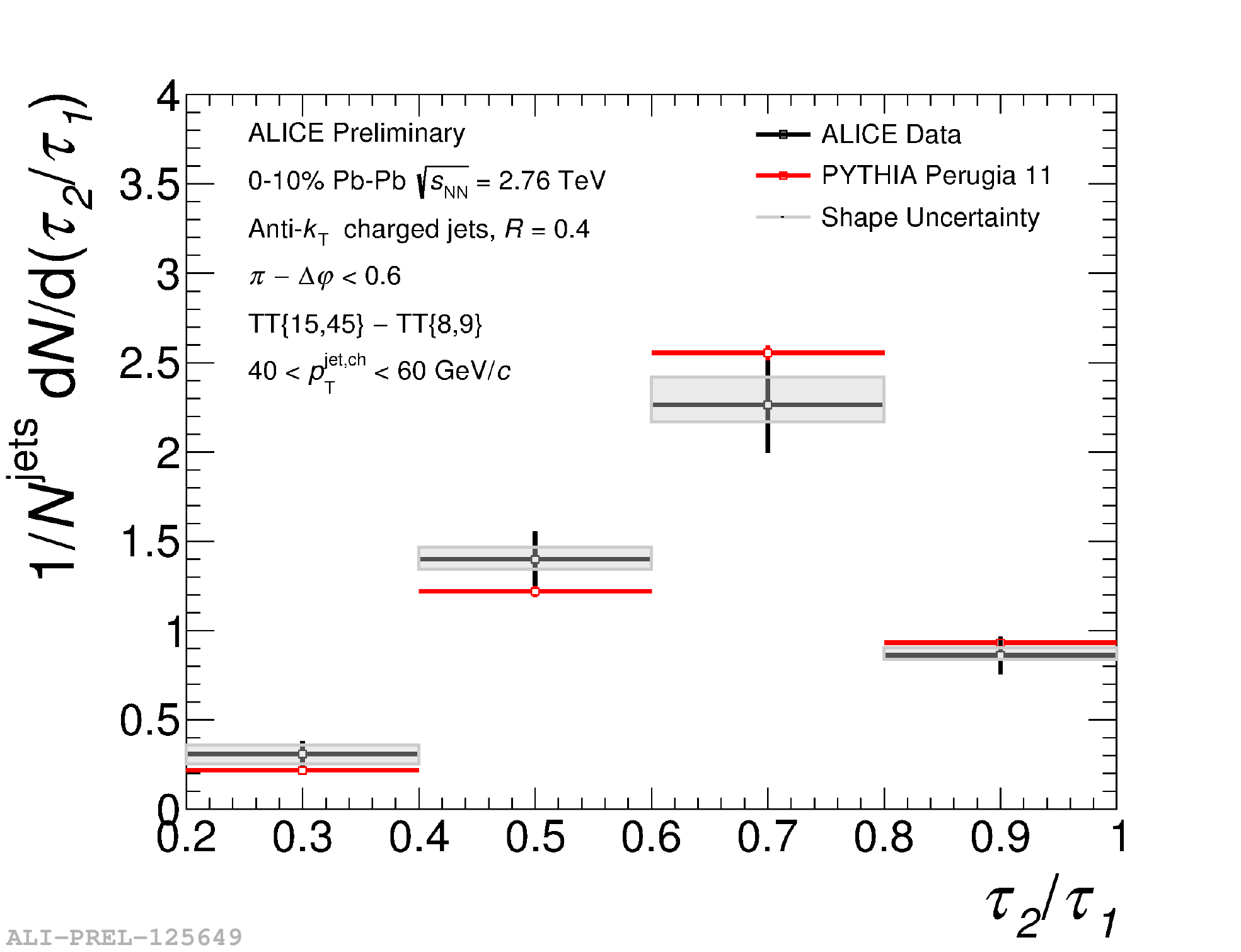}
\centering
\caption{Left: Radial moment $g$ for Pb--Pb collisions hinting to the collimation of charged jets with $R=0.2$. Right: N-subjettiness $\tau_2/\tau_1$ measuring the 2-prongedness of charged jets with $R=0.4$. \protect\cite{JetShapes,Nsubjettiness}}
\label{fig:Panel1}
\end{figure}

\section{Results}
\label{sec:Results}

In Fig.~\ref{fig:Panel1}, radial moment $g$ and $N$-subjettiness with $N=2$ are presented for the 10\% most central Pb--Pb collisions at $\snn = 2.76$ TeV.\,\cite{JetShapes,Nsubjettiness} Both quantities shed light on the jet substructure and its changes in a medium. While the radial moment is a measure for the collimation of the jet momentum, the $N$-subjettiness observable $\tau_2/\tau_1$ indicates how 2-pronged the jet is. Due to color coherence effects, the probability for 2-pronged jets could be different in Pb--Pb compared to pp.
The measurement of $g$ indicates that the jet core is more collimated in the medium, compared to PYTHIA\,\cite{PYTHIA6}. Also the medium-modified jets in JEWEL\,\cite{JEWEL} qualitatively agree with the data. 
Within the uncertainties, $\tau_2/\tau_1$ indicates no strong quenching effect on the jet substructure.\\

\begin{figure}[htp]
\includegraphics[width=0.98\textwidth]{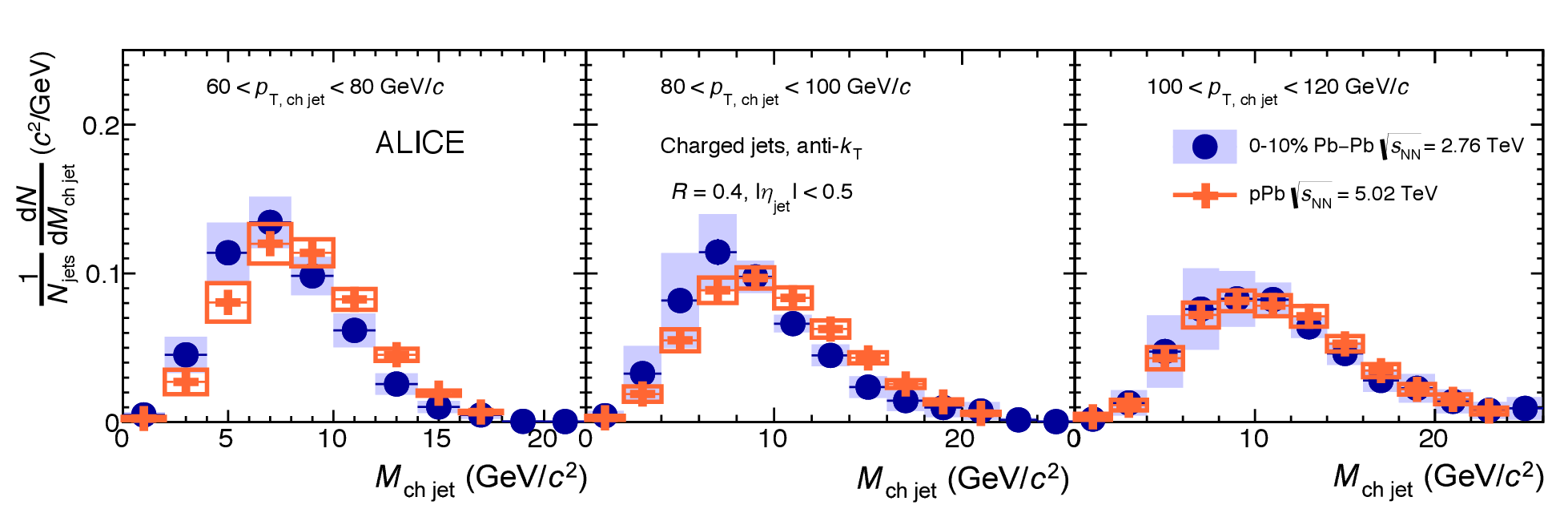}
\centering
\caption{Jet mass for p--Pb and the 10\% most central Pb--Pb collisions for charged jets with $R=0.4$. \protect\cite{JetMass}}
\label{fig:Panel2}
\end{figure}

Figure~\ref{fig:Panel2} shows another quantity connected to the jet substructure: The jet mass in Pb--Pb, compared to p--Pb collisions.\,\cite{JetMass} We observe an indication for a shift in jet mass to lower values comparing Pb--Pb and p--Pb collisions. This difference cannot be explained by the difference in $\snn$ of both measurements and could hint to a higher collimation of the jet core in Pb--Pb.

\begin{figure}[htp]
\includegraphics[width=0.39\textwidth]{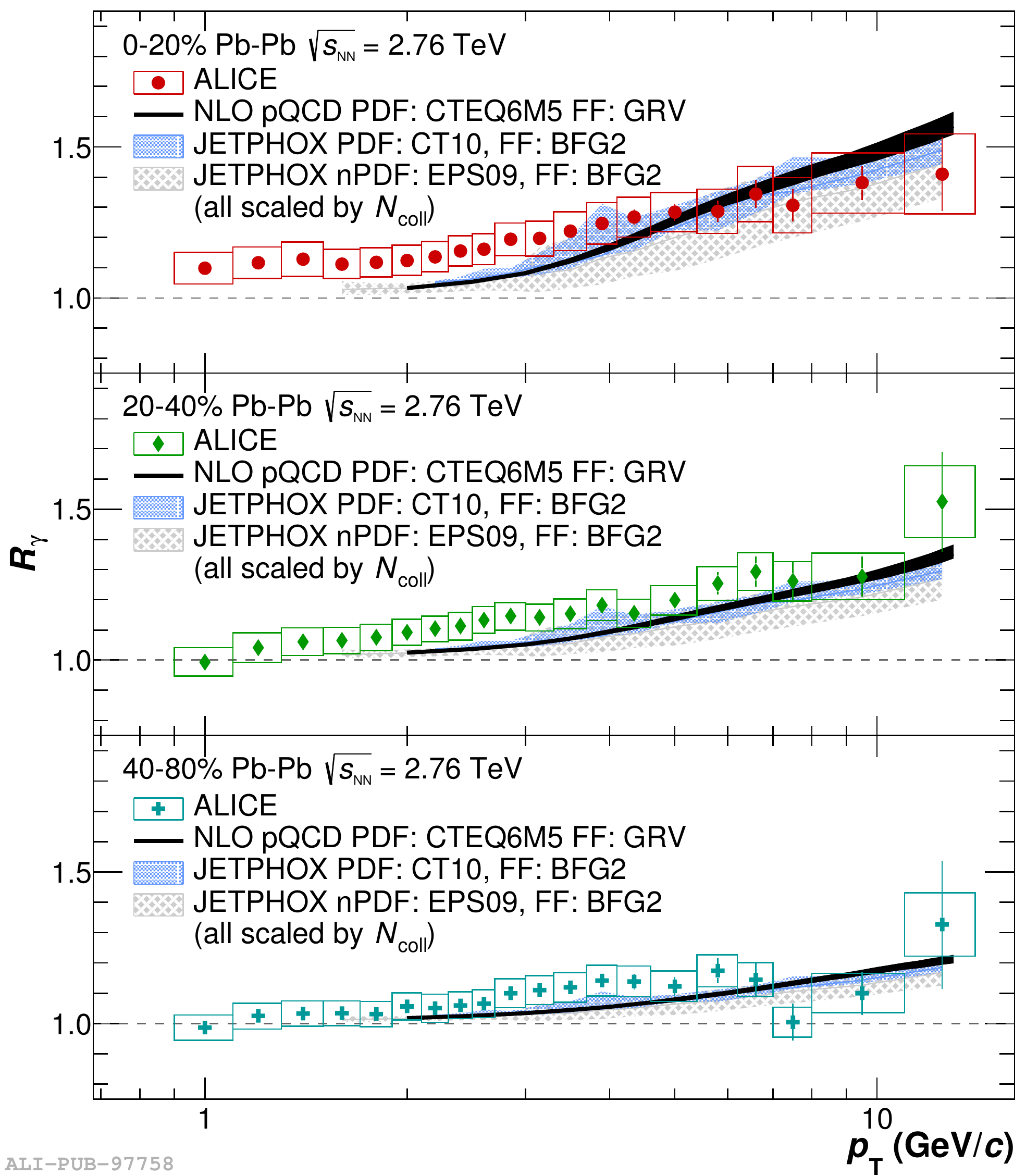}
\includegraphics[width=0.39\textwidth]{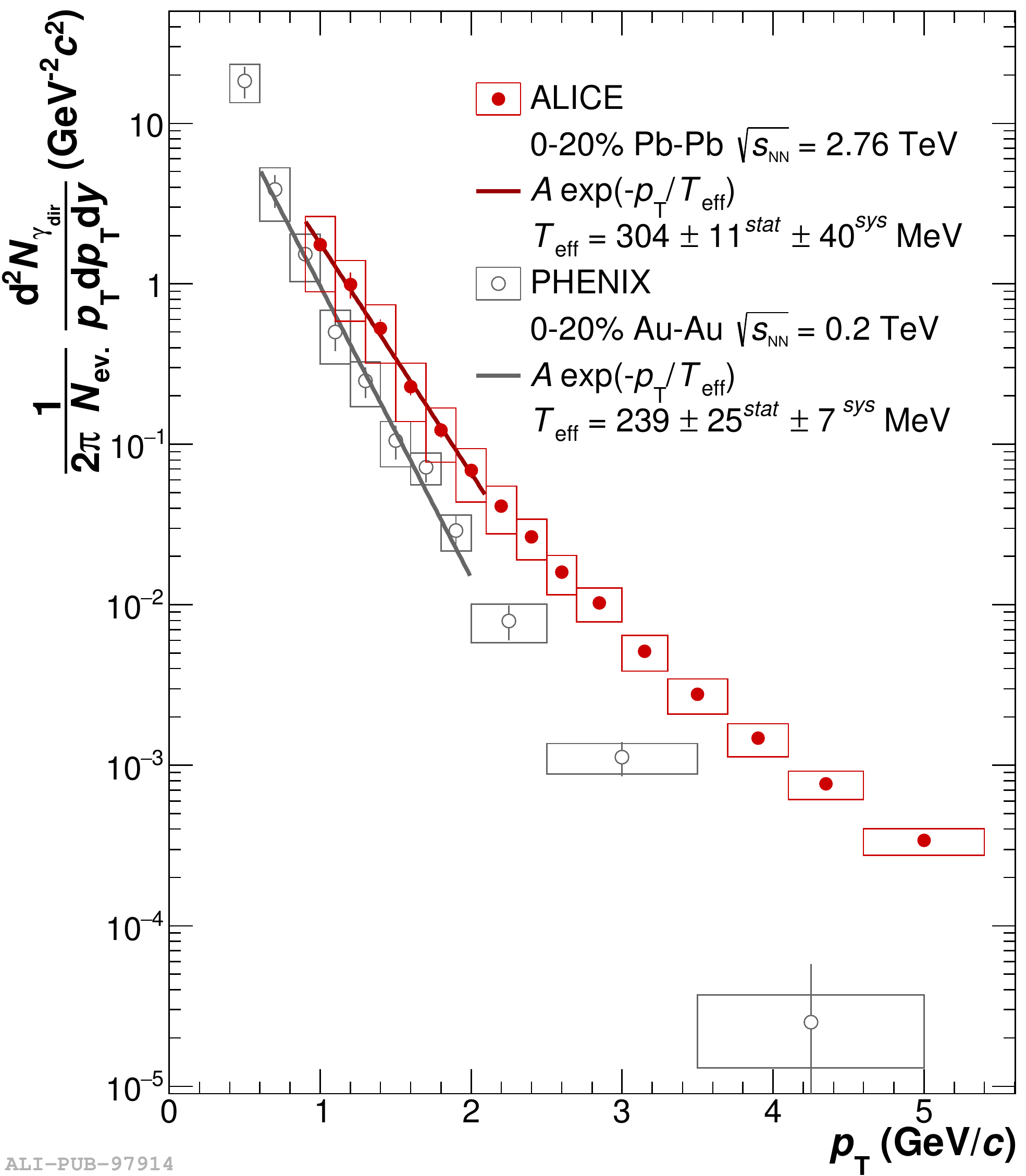}
\centering
\caption{Left: Direct photon double ratio for Pb--Pb collisions at $\snn = 2.76$ TeV for several centrality bins compared to pQCD calculations. Right: Direct photon spectra for Pb--Pb collisions at $\snn = 2.76$ TeV measured by ALICE compared to PHENIX data. \protect\cite{PhotonPaper,PhotonPaperSup}}
\label{fig:Panel3}
\end{figure}

\begin{figure}[htp]
\includegraphics[width=0.39\textwidth]{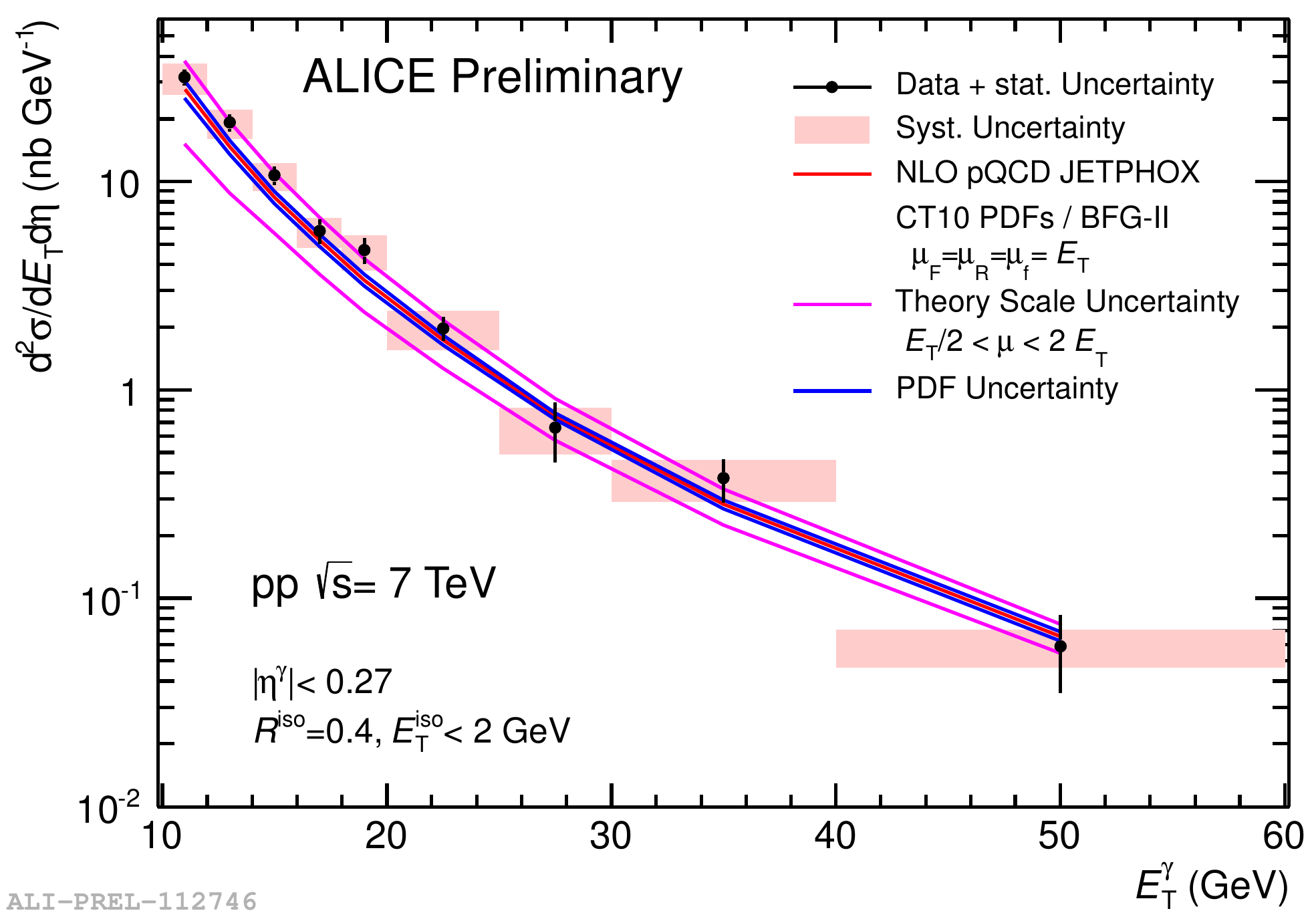}
\includegraphics[width=0.39\textwidth]{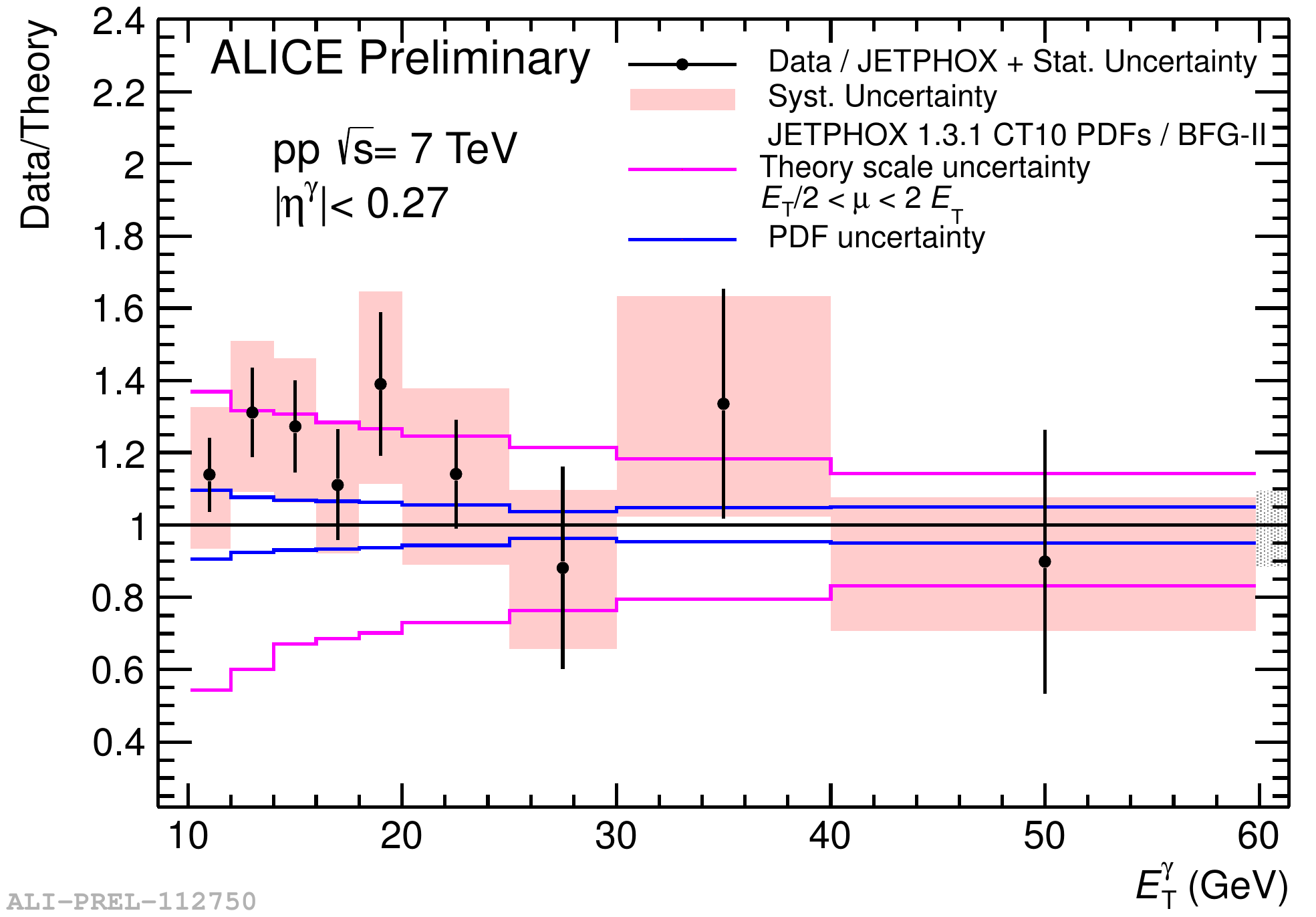}
\centering
\caption{Left: Isolated photon cross section for pp collisions. Right: Isolated photon theory comparison.} 
\label{fig:Panel4}
\end{figure}

Photon measurements are depicted in Figs.~\ref{fig:Panel3} and \ref{fig:Panel4}. In Fig.~\ref{fig:Panel3}, direct photon production at $\snn = 2.76$ TeV in Pb--Pb collision is presented. The left plot shows the direct photon double ratio $R_\gamma$, which already indicates an enhancement at low $\pT$ for more central collisions. This enhancement can be interpreted as thermal photons. In the right plot, the direct photon spectrum with its excess at low $\pT$ is shown. A comparison to PHENIX data reveals the higher effective temperature of the medium at the LHC compared to RHIC.\,\cite{PhotonPaper,PhotonPaperSup}

A measurement of isolated photons in pp collisions at $\sqrt{s} = 7$ TeV is shown in Fig.~\ref{fig:Panel4}. A good agreement of the data with pQCD calculations is shown in the right plot. This measurement is a first step to an application in Pb--Pb collisions. In addition, isolated photon measurements might be of particular interest for photon-jet events in Pb--Pb collisions.

\section{Summary}
\label{sec:Summary}

The jet shapes, in particular the radial momentum, are sensitive to the interaction with the medium produced in nuclear collisions. In Pb--Pb collisions, there is an indication for a stronger momentum collimation. In addition, the 2-pronged subjet structure is not strongly affected.

The measurement of direct photons in Pb--Pb collisions shows an excess at low transverse momenta hinting to the presence of thermal photons. The thermal radiation indicates a higher effective temperature at the LHC than at RHIC.



\end{document}